\documentclass[12pt]{article}
\usepackage{amsmath,amssymb}
\usepackage{graphicx,color}
\numberwithin{equation}{section}
\usepackage{cite}
\usepackage{bm}
\usepackage{dcolumn}
\usepackage{xcolor}

\usepackage{cancel} 
\usepackage[final]{changes}

\newcommand{\be}{\begin{equation}}
\newcommand{\bea}{\begin{eqnarray}}
\newcommand{\eea}{\end{eqnarray}}
\newcommand{\ba}{\begin{array}}
\newcommand{\ea}{\end{array}}
\newcommand{\ee}{\end{equation}}

\expandafter\ifx\csname mathbbm\endcsname\relax

\else

\fi \textheight 22cm \textwidth 15cm \topmargin 1mm \oddsidemargin
5mm \evensidemargin 5mm

\begin{document}
\begin{titlepage}
\hfill \vbox{
    \halign{#\hfil         \cr
         \cr
                      } 
      }  
\vspace*{20mm}
\begin{center}
{\Large {\bf Applying Constraints from Outside \deleted{of} the System}\\
}

\vspace*{15mm} \vspace*{1mm} {Amin Akhavan}

\vspace*{.4cm}

{\it  School of Particles and Accelerators, Institute for Research in Fundamental Sciences (IPM)\\
P.O. Box 19395-5531, Tehran, Iran\\$email:amin_- akhavan@ipm.ir$ }

\vspace*{2cm}

\end{center}

\begin{abstract}
\deleted{We derive the representation of the external constraints imposed on the system within the generating functional. These constraints are applied to both the degrees of freedom and their time derivatives,
and the generating functional is expressed as a functional integral over the degrees of freedom. To achieve this, we employ the path integral method for handling second-class constraints, which we have briefly reviewed.
We use the obtained result to derive the generating functional for matter vector fields constrained to the spin-one state. However, achieving the desired result for a system with a singular action requires certain sufficient conditions that we will obtain.} 

\added{We investigate the role of external constraints in quantum field theory using the path integral formalism. We begin by reviewing the quantization of constrained systems and extend the analysis to cases where constraints are added to the action via auxiliary fields. These constraints involve both the degrees of freedom and their time derivatives. Using the resulting framework, we derive the generating functional for matter vector fields constrained to the spin-one state. However, for systems with singular actions, preserving the classical form of constraints at the quantum level requires conditions, and we provide examples of sufficient conditions. We also present an example where the classical form of constraints is not preserved, involving a singular action with an antisymmetric tensor field leading to a strength-field interpretation.}
\\\\ Keywords: External Constraints \\PACS numbers: 03.70.+k
\end{abstract}

\vspace{2cm}

\end{titlepage}

\section{Introduction}

\,\,\,\,\,\,\,\, \deleted{We previously considered the presence of constraints outside the equations of motion within the functional path integral \cancel{\cite{akh}}. We wrote such constraints as functionals of the degrees of freedom. Using Dirac delta functionals of constraints, 
we can have sources for constrained degrees of freedom and define effective actions and spontaneous symmetry breaking processes\cancel{\cite{fuk, cole},\cite{symanz}}. We can use the Lagrange multipliers such as sources, which 
fix the background state so that it is not a vacuum\cancel{\cite{symanz,akha}}. Also, we can use Lagrange multipliers such as degrees of freedom in the action, to fix the background field so that the effective potential is not symmetric.}

\deleted{Delta functionals in the functional integrals fix the degrees of freedom. We have gauge symmetries and constraints in the systems with the gauge fields so that we can fix the gauges\cancel{\cite{dir, fad}}. However, fixing degrees of freedom of the systems 
without gauge symmetry indicates constraints incorporated from outside the system\cancel{\cite{dira}}. Such a fixing represents a projection onto eigenstates, where the fixed degrees of freedom correspond to their eigenvalues.This projection is a process
that occurs in the unity of a macroscopic system and a microscopic one.  In other words, when such a fixing occurs, certain quantities become determined and collectively form macroscopic systems.}

\deleted{Let us consider a measurement as a mechanism in which some quantities can be accurately measured, such as spontaneously measured quantities. In this case, we will have two types of quantum and classical degrees of freedom. 
The classical types are completely out of the functional integral. Although the quantum variables remain within the functional integral, they must be constrained to ensure compatibility with the classical quantities. These constraints typically appear in the form of Dirac delta functionals.
For example, the derivatives of the scalar fields have classical backgrounds that allow us to have massive gravity\cancel{\cite{muk}}. Thus, we can construct theories of classical backgrounds, with quantum perturbations superimposed upon them.}

\deleted{A specific method exists for quantizing actions with first- or second-class constraints\cancel{\cite{fadi,senj}}. This method was originally designed to solve the problem of gauge fixing. But we can use it for actions with Lagrange multipliers. 
The key question is: does the use of the Lagrange multipliers and Dirac quantization enable the presence of delta functionals of the constraints in the path integrals?
(In other words, if we impose a constraint on the classical behavior of a theory, will the same constraint also apply to its quantum vacuum?)}

\deleted{Section 2 reviews constrained systems described by actions with quadratic kinetic terms. These systems have two samples of constraints: first-class and second-class. We examine the role of these constraints in the functional path integral. In this paper, we use Senjanovic's method\cancel{\cite{senj}}. In section 3, we derive the generating functional for a nonsingular action with external constraints imposed on it, considering constraints that involve time derivatives of the fields. The inclusion of time derivatives is crucial for achieving the desired result. In Section 4, we apply our method to derive the generating functional for a 4-vector matter field theory, incorporating an external constraint that enforces the spin of the particle to be one.
We performed similar calculations for singular actions in the fifth and sixth sections, obtaining sufficient conditions to achieve results analogous to those in the third section. In the last section, we talked about the interpretation 
that can be made of the external constraints applied to the theories.}

\added{In quantum field theory, constraints play a fundamental role in shaping the behavior of fields and their interactions. Traditional approaches typically incorporate constraints as conditions arising from the equations of motion or gauge symmetries }\cite{dir,fad,fadi,H}. \added{However, an alternative perspective considers the imposition of external constraints that influence the theory from outside its intrinsic dynamics. Such constraints can be formulated using delta functionals in the path integral, ensuring that certain degrees of freedom remain fixed-thereby creating a background state that deviates from the conventional vacuum state}\cite{fuk,muk,akh,taro,toms}.

\added{A key motivation for introducing external constraints is to control background fields in a way that distinguishes between classical and quantum degrees of freedom. In particular, imposing constraints through delta functionals in the path integral formulation projects the theory onto subspaces where specific quantum field configurations are confined, while classical quantities are externally fixed to define those subspaces. This approach is reminiscent of spontaneous symmetry breaking}\cite{cole,symanz}\added{,and is conceptually related to the background field method}\cite{abb}\added{, yet it extends beyond conventional treatments where expectation values of fields are merely shifted by a constant.  In this paper, we aim to impose dynamical background structures, allowing classical fields to co-exist with quantum fluctuations in a systematic manner.}

\added{We introduce a novel framework for quantizing systems with constraints that depend on time derivatives of the fields, building on earlier investigations of constrained systems} \cite{fuk,dira,senj,akha}. \added{These constraints are implemented via Lagrange multipliers, treated as auxiliary degrees of freedom within the action. Integrating over these degrees of freedom in the path integral can, in some cases, lead to delta functionals that enforce the constraints.  In general, our constraints depend on both the fields and their time derivatives, i.e., $f(q,\dot{q})=0$. This raises the important question of whether such classical constraints are preserved in the quantum level. Additionally, we examine how the partition function is modified by the presence of such constraints.}

\added{A centeral innovation of our work is the systematic treatment of constraints involving first-order time derivatives as classical background fields.  These backgrounds govern the quantum dynamics while remaining unaffected by quantum fluctuations. This perspective enables quantum fields to evolve under prescribed classical dynamics, maintaining consistent quantization conditions. The framework naturally accommodates scenarios where background field equations are imposed dynamically.}

\added{In this paper, we systematically study the consequences of imposing external constraints within the path integral formalism. We begin in Section 2 by reviewing Senjanovic's method of  constrained systems with quadratic kinetic terms, discussing both first-class and second-class constraints and their impact on quantum dynamics}\cite{senj}. \added{In Section 3, we derive the generating functional for a constrained system in which time derivatives of the fields play a crucial role. Section 4 applies this methodology to a vector field theory, demonstrating how an external constraint can enforce the particle’s spin to be one. Section 5 extends the analysis to singular actions, identifying sufficient conditions under which the classical form of constraints is preserved at the quantum level. Section 6 discusses further generalizations and alternative formulations for the sufficient conditions. Section 7 presents an explicit example involving a singular action with an antisymmetric tensor field, where the classical form of constraints is not preserved, leading instead to a strength-field interpretation. Finally, in Section 8, we interpret the role of these constraints in defining a macroscopic background and discuss how the constraint forces can be eliminated, leading to a purely Hamiltonian description of the constrained system.}

\added{This perspective provides a possible framework for formulating quantum field theories on structured classical backgrounds.In the context of quantum gravity, understanding the emergence of classical spacetime from a quantum theory remains a central challenge. Our future work suggests a potential mechanism for such emergence.}

\section{Second-class\added{ and first-class} constraints}
We aim to analyze quantum fields with quadratic Lagrangians that take a simple form in the path integral method:
\be\label{2.1}
L=\frac{1}{2}A_{ij}(q)\dot{q}_{i}\dot{q}_{j}+B_{i}(q)\dot{q}_{i}-V(q),~~~~~~~~~i=1,...,N.
\ee
where the discrete spatial coordinates are included in the index $i$. 
If $\det A_{ij}=0$, \deleted{we call them singular, }\added{we classify the system as singular, indicating} that some canonical momenta are dependent. Consider a 2N-dimensional phase space of degrees of freedom and their canonical momenta for the system. \deleted{Since in the}\added{In the case of} singular Lagrangians, \added{since} some of the momenta are not independent, there exist\deleted{some} constraints that\deleted{limit} \added{restrict} the system to a subspace of the phase space{, as discussed by Senjanovic \cite{senj}}. We can \deleted{show}\added{express} the dependencies of the momenta with these primary constraints:
\be\label{2.2}
\psi_{\alpha}(q_{i},p_{i})=0,~~~~~~~\alpha=1,...,M_{1},
\ee
where the subspace has $2N-M_{1}$ dimensions. To find all $\dot{q}_{i}$ as functions of  $p_{i}$, we consider:
\be\label{2.3}
p_{i}=A_{ij}(q)\dot{q}_{j}+B_{i}(q), 
\ee
since the momenta are dependent, we define independent quantities $x_{\alpha}$ in which we can write $\dot{q}_{i}$ as functions of $x_{\alpha}, p_{i}, q_{i}$.
Now we can define a Hamiltonian like this:
\be\label{2.4}
H=\dot{q}_{i}(x_{\alpha}, q_{i},p_{i})p_{i}-L(q_{i},\dot{q}_{i}(x_{\alpha}, q_{i},p_{i})).
\ee
By applying the definition, it is clear that: 
\be\label{2.5}
\frac{\partial H}{\partial x_{\alpha}}=\frac{\partial \dot{q}_{i}}{\partial x_{\alpha}}p_{i}-\frac{\partial L}{\partial \dot{q}_{i}}\frac{\partial \dot{q}_{i}}{\partial x_{\alpha}}=0,
\ee
such that the Hamiltonian is only function of $q$, $p$ in the subspace. \deleted{Also, we can have:}\added{However, to derive the Hamiltonian equations, the Hamiltonian must be formulated over the entire phase space prior to the imposition of any constraints:}
\bea\label{2.6}
\dot{q}_{i}=\frac{\partial H}{\partial p_{i}}~~\nonumber\\
\dot{p}_{i}=-\frac{\partial H}{\partial q_{i}}.
\eea
\deleted{Although there are different Hamiltonians with different functionality of momentums,
they are all the same in the subspace.}\added{Although different Hamiltonians can be formulated with different dependencies on the momenta, they all lead to equivalent dynamics in the subspace defined by the primary constraints. More fundamentally, extending to the full phase space introduces redundant degrees of freedom, and only by restricting the resulting Hamiltonian equations to the constraint surface do we obtain a unique physical description.} \\ 

\deleted{Apart from}\added{In addition to} the \added{primary} constraints \deleted{mentioned}\added{given} in equation \eqref{2.2}, \deleted{we can obtain new}\added{furhter} constraints \added{can arise} from the equations of motion. In fact, for each primary constraint, \deleted{one can find a non-dynamic equation of motion that is not function of}\added{there exists a corresponding equation of motion that is non-dynamical—i.e., it does not involve} $\ddot{q}_{i}$.
In these non-dynamic\added{al} equations,\deleted{we can write} $\dot{q}_{i}$\added{ can be expressed in terms} \deleted{as functions} of $x_{\alpha}, p_{i}, q_{i}$. \deleted{There is no additional constraint if all $x_{\alpha}$ can be obtained in these equations.}\added{ If all the $x_{\alpha}$ can be determined from these relations, no additional constraints emerge.} \deleted{But as many $x_{\alpha}$ remain undetermined, the derived equations will only be functions of  $p_{i}, q_{i}$, and they will be 
referred to as secondary constraints. We show them as follows:}\added{However, when some of the $x_{\alpha}$ remain undetermined, the resulting equations become independent of those $x_{\alpha}$ and involve only $p_{i}$ and $q_{i}$; these are identified as secondary constraints, which we denote as follows:}
\be\label{2.6a}
\phi_{\sigma}(q_{i},p_{i})=0,~~~~~~~\sigma=1,...,M_{2},
\ee 
thus the phase space will be reduced to a new subspace with $2N-(M_{1}+M_{2})$ dimensions.\\

As \deleted{the} \added{a} first necessary condition, if $M_{1}+M_{2}$ is \deleted{an} even\deleted{number}, \deleted{we can quantize the system in this subspace.}\added{the system can be quantized within this subspace.} \deleted{First, we have to define new coordinates for the phase space}\added{To proceed, we introduce new canonical coordinates} ($q^{*}_{i},p^{*}_{i})$ \deleted{we can write them in two parts:}\added{ for the phase space, which can be partitioned into two subsets:}
 
\be\label{2.6e}
(q^{*}_{a},p^{*}_{a}) ~~~~~~~~~~~~~a=1,...,\frac{M_{1}+M_{2}}{2}~~~~~~~
\ee
\be\label{2.6f}
(q^{*}_{A},p^{*}_{A}) ~~~~~~~~~~~~~A=1,..., N-\frac{M_{1}+M_{2}}{2},
\ee
\deleted{which}\added{where} 
$q^{*}_{a}=p^{*}_{a}=0$ \deleted{shows}\added{defines} \deleted{the points in} the subspace, and $q^{*}_{A},p^{*}_{A}$ are the coordinates \deleted{over}\added{on} the subspace. The \deleted{important}\added{crucially} point is that these new coordinates must be canonical like the 
original coordinates. It means:
\be\label{2.6a}
 \{q^{*}_{i},q^{*}_{j}\}=0, ~\{p^{*}_{i},p^{*}_{j}\}=0,~ \{q^{*}_{i},p^{*}_{j}\}=\delta_{ij}.
\ee
In this case, if
\be\label{2.6b}
A^{*}(q^{*},p^{*})=A(q,p),~~~~~~B^{*}(q^{*},p^{*})=B(q,p),
\ee
then, using the chain rule, the following equation can be obtained:
\be\label{2.6c}
\{A,B\}=\frac{\partial A^{*}}{\partial q^{*}_{i}}\frac{\partial B^{*}}{\partial p^{*}_{i}}-\frac{\partial A^{*}}{\partial p^{*}_{i}}\frac{\partial B^{*}}{\partial q^{*}_{i}}=\{A^{*},B^{*}\}^*,
\ee
and therefore we get the equations \eqref{2.6} in the new coordinates, if we have:
\be\label{2.6d}
H^{*}(q^{*},p^{*})=H(q,p).
\ee  
In continuation, we can define a functional path integral in the subspace:
\be\label{2.7}
\int \mathcal{D}q^{*}_{A}\mathcal{D}p^{*}_{A}e^{i\int dt \dot{q}^{*}_{A}p^{*}_{A}-H^{*}(q^{*}_{a}=0,p^{*}_{a}=0,q^{*}_{A},p^{*}_{A})},
\ee
that we can write it like:
\be\label{2.8}
\int \mathcal{D}q^{*}_{i}\mathcal{D}p^{*}_{i}\delta(q^{*}_{a})\delta(p^{*}_{a})e^{i\int dt \dot{q}^{*}_{i}p^{*}_{i}-H^{*}(q^{*}_{i},p^{*}_{i})}.
\ee
Before continuing the above equation, we need to explain the canonical transformation. We have to consider an infinitesimal canonical transformation;
\be\label{2.9}
q^{'}_{i}=q_{i}+\delta q_{i}, ~~~~~p^{'}_{i}=p_{i}+\delta p_{i},
\ee
such that, from equations like\eqref{2.6a} we will have:
\be\label{2.19}
\{q_{i},\delta q_{j}\}+\{\delta q_{i},q_{j}\}=0, ~\{p_{i},\delta p_{j}\}+\{\delta p_{i},p_{j}\}=0,~\{q_{i},\delta p_{j}\}+\{\delta q_{i},p_{j}\}=0,
\ee
which is written as follows:
\be\label{2.20}
\frac{\partial \delta q_{j}}{\partial p_{i}}-\frac{\partial \delta q_{i}}{\partial p_{j}}=0,~~\frac{\partial \delta p_{j}}{\partial q_{i}}-\frac{\partial \delta p_{i}}{\partial q_{j}}=0,~~\frac{\partial \delta p_{j}}{\partial p_{i}}+\frac{\partial \delta q_{i}}{\partial q_{j}}=0.
\ee
The latter equations are the same as the zero curl of a vector. So we can write such a vector as a gradient of a scalar:
\be\label{2.21}
\delta q_{i}=\epsilon\frac{\partial F(q,p)}{\partial p_{i}},~~ \delta p_{i}=-\epsilon\frac{\partial F(q,p)}{\partial q_{i}}.
\ee
Using these equations we can have:
\be
\delta\int dt\dot{q}_{i}p_{i}=\int dt \frac{d}{dt}(\delta q_{i}p_{i})-\delta q_{i}\dot{p}_{i}+\delta p_{i}\dot{q}_{i}\nonumber\\
\ee
\be\label{2.22}
=\bigg(\epsilon p_{i}\frac{\partial F}{\partial p_{i}}-\epsilon F\bigg)^{+\infty}_{-\infty}=0.~~~~~~~~~~~
\ee
And also,
\be\label{2.22a}
\delta(\mathcal{D}q_{i}\mathcal{D}p_{i})=tr\begin{bmatrix}\frac{\partial\delta q_{i}}{\partial q_{j}}&\frac{\partial\delta q_{i}}{\partial p_{j}}\\
\frac{\partial\delta p_{i}}{\partial q_{j}}&\frac{\partial\delta p_{i}}{\partial p_{j}}\end{bmatrix}\mathcal{D}q_{i}\mathcal{D}p_{i}\nonumber\\
\ee
\be\label{2.23}
=(\frac{\partial\delta q_{i}}{\partial q_{i}}+\frac{\partial\delta p_{i}}{\deleted{\cancel{\partial q_{i}}} \added{\partial p_{i}}})\mathcal{D}q_{i}\mathcal{D}p_{i}=0.
\ee
By puting the results obtained by equations \eqref{2.6d},\eqref{2.22} and \eqref{2.23} in equation \eqref{2.8}, we will have:  
\be\label{2.24}
\int \mathcal{D}q_{i}\mathcal{D}p_{i}\delta(q^{*}_{\alpha})\delta(p^{*}_{\alpha})e^{i\int dt \dot{q}_{i}p_{i}-H(q_{i},p_{i})}.
\ee
We define the set of constraints as a vector like $\Phi_{I}=(\psi_{\alpha}, \phi_{\sigma})$ and the coordinates vector like $\rho^*_{I}=(q^*_{a},p^*_{a})$. 
Now using the Jacobian method, we can write delta functions in equations \eqref{2.24} as below,
\be\label{2.25}
\delta(q^*_{a})\delta(p^*_{a})=\delta(\rho^*_{I})=\det(\frac{\partial\Phi_{I}}{\partial\rho^*_{J}})\delta(\Phi_{I}).
\ee
For infinitesimal $\rho^*_{I}$, we can have $\Phi_{I}(\rho^*_{J})=\Lambda_{IJ}\rho^*_{J}+...$ and therefore we have:
\be\label{2.27}
\delta(\rho^*_{I})=(\det(\Lambda_{IJ})+...)\delta(\Phi_{I}).
\ee
In addition, using equation \eqref{2.6c}, we have:
\be\label{2.28}
\{\Phi_{I},\Phi_{J}\}=\{\Lambda_{IK}\rho^*_{K}+...,\Lambda_{JL}\rho^*_{L}+...\}^*=\Lambda_{IK}\{\rho^*_{K},\rho^*_{L}\}^*\tilde{\Lambda}_{LJ}+...,
\ee
thus,
\be\label{2.29}
\det\{\Phi_{I},\Phi_{J}\}=(\det(\Lambda_{IJ}))^2\det\{\rho^*_{K},\rho^*_{L}\}^*+...=(\det(\Lambda_{IJ}))^2+...,
\ee
and
\be\label{2.30}
\delta(\rho^*_{I})=((\det\{\Phi_{I},\Phi_{J}\})^{\frac{1}{2}}+...)\delta(\Phi_{I}).
\ee
In the presence of delta functions, we can skip the perturbational sentences. By inserting equation \eqref{2.30} into equation \eqref{2.24}, we have:
\be\label{2.31}
\int \mathcal{D}q_{i}\mathcal{D}p_{i}(\det\{\Phi_{I},\Phi_{J}\})^{\frac{1}{2}}\delta(\Phi_{I})e^{i\int dt \dot{q}_{i}p_{i}-H(q_{i},p_{i})}.
\ee
\deleted{Now, we can express}\added{We can now state} the necessary condition as $\det\{\Phi_{I},\Phi_{J}\}\neq0$. This condition \deleted{also encompasses the requirement of evenness}\added{implicitly ensures that the total number of constraints is even}, since the determinant of an odd antisymmetric matrix \deleted{is zero}\added{necessarily vanishes}.
There are simple examples whose primary constraints are $p_{\alpha}=0$, in which case:
\be\label{2.32}
\det\{\Phi_{I},\Phi_{J}\}=\begin{vmatrix}0&\{p_{\alpha},\phi_{\beta}\}\\-\{p_{\alpha},\phi_{\beta}\}&\{\phi_{\alpha},\phi_{\beta}\}\end{vmatrix}=(\det{\{p_{\alpha},\phi_{\beta}\}})^2,
\ee
and the path integral is given by:
\be\label{2.33}
\int \mathcal{D}q_{i}\mathcal{D}p_{i}\det(\frac{\partial\phi_{\beta}}{\partial q_{\alpha}})\delta(p_{\alpha})\delta(\phi_{\alpha})e^{i\int dt \dot{q}_{i}p_{i}-H(q_{i},p_{i})}.
\ee

So far, we have proposed the necessary condition, which we could show as a non-zero determinant of the Poisson bracket matrix of constraints. However, if the value of the determinant becomes zero,
the relationship between the constraints and the subspace in which the quantization process takes place should be analyzed more carefully.  \deleted{If the determinant is zero, 
the Poisson bracket matrix can be written by a linear transformation of the constraints so that some rows are completely zero. With the same number of zero rows, some constraints commute with all other constraints. We refer to the set of these commutable constraints as a first-class set. We show these first-class constraints as}\added{In such cases, the Poisson bracket matrix can be brought—via a linear redefinition of the constraints—into a form where some rows (and corresponding columns) are identically zero. The number of these zero rows corresponds to the number of constraints that commute with all others. These mutually commuting constraints are referred to as first-class constraints. We denote them by} $\Psi_{\mu}(q_{i},p_{i})=0$, \deleted{that we have}\added{satisfying}:
\be\label{2.34}
\{\Psi_{\mu},\Psi_{\nu}\}=0, ~~~~~~~~~~~\{\Psi_{\mu},\Gamma_{m}\}=0,
\ee
that other constraints $\Gamma_{m}$, are second-class.
Now we define new canonical coordinates $(q^*_{i},p^*_{i})$ in the form of duals like:
\be\label{2.35}
(q^{*}_{a},p^{*}_{a}), ~~~~~~~~(q^{*}_{A},p^{*}_{A}), ~~~~~~~~~(q^{*}_{\mu},p^{*}_{\mu}).
\ee
In these canonical coordinates, the constraints are:
\be\label{2.36} 
p^{*}_{\mu}=q^{*}_{a}=p^{*}_{a}=0.
\ee
Compared to equation \eqref{2.29}, the determinant of the bracket of these constraints has to be zero, and like equations \eqref{2.34} we have:
\be\label{2.37}
\{p^*_{\mu},q^*_{a}\}=0, ~~~~~~~~~~~\{p^*_{\mu},p^*_{a}\}=0.
\ee
And the coordinates over the subspace are:
\be\label{2.38}
(q^{*}_{\mu},q^{*}_{A},p^{*}_{A}).
\ee
Since the constraints must be true for all times, then in the subspace we have:
\be\label{2.39}
\{H^*,P^*_{\mu}\}=0,
\ee
and since $q^*_{\mu}$ is canonical couple of $P^*_{\mu}$ we can write:
\be\label{2.40}
\frac{\partial H^*}{\partial q^*_{\mu}}=\{H^*,P^*_{\mu}\}=0.
\ee
The recent equation shows a symmetry for the Hamiltonian called gauge symmetry. Since the Hamiltonian is not a function of $q^*_{\mu}$, and also $p^*_{\mu}$ are not coordinates in the subspace, we can define the functional path integral as follows:
\be\label{2.41}
\int \mathcal{D}q^{*}_{A}\mathcal{D}p^{*}_{A}e^{i\int dt \dot{q}^{*}_{A}p^{*}_{A}-H^{*}(p^*_{\mu}=0,q^{*}_{a}=0,p^{*}_{a}=0,q^{*}_{A},p^{*}_{A})},  
\ee
and so,
\be\label{2.41}
\int \mathcal{D}q_{i}\mathcal{D}p_{i}\delta(q^{*}_{\mu})\delta(p^{*}_{\mu})\delta(q^{*}_{a})\delta(p^{*}_{a})e^{i\int dt \dot{q}_{i}p_{i}-H(q_{i},p_{i})}.  
\ee
We have $q^{*}_{\mu}=0$  as new constraints in the recent equations. These constraints fix the gauge symmetry which we call gauge fixing constraints. In the original coordinates of the phase space, we assign the role of the gauge fixing process to $X_{\mu}(q_{i},p_{i})=0$. 
Compared to equations \eqref{2.24} to \eqref{2.31}, we define $\rho_{I}=(q^{*}_{\mu},p^{*}_{\mu}, q^{*}_{a},p^{*}_{a})$ and $\Phi_{I}=(\Psi_{\mu},X_{\mu},\Gamma_{M})$. In this definition, we can obtain,
\be
\det\{\Phi_{I},\Phi_{J}\}=\begin{vmatrix}0&\{\Psi_{\mu},X_{\nu}\}&0\\ \{X_{\mu},\Psi_{\nu}\}&\{X_{\mu},X_{\nu}\}& \{X_{\mu},\Gamma_{M}\}\\ 0&\{\Gamma_{M},X_{\mu}\}&\{\Gamma_{M},\Gamma_{N}\}\end{vmatrix}\nonumber\\
\ee
\be\label{2.42}
=(\det\{\Psi_{\mu},X_{\nu}\})^2\det\{\Gamma_{M},\Gamma_{N}\}.
\ee
By substituting the obtained result for the determinant, equation \eqref{2.31} becomes as follows: 
\be\label{2.43}
\int \mathcal{D}q_{i}\mathcal{D}p_{i}\det\{\Psi_{\mu},X_{\nu}\}(\det\{\Gamma_{M},\Gamma_{N}\})^{\frac{1}{2}}\delta(\Phi_{I})e^{i\int dt \dot{q}_{i}p_{i}-H(q_{i},p_{i})}.
\ee
Now, we can see the necessary condition $\det\{\Psi_{\mu},X_{\deleted{\cancel{\mu}}\added{\nu}}\}\neq 0$, which we use to determine $X_{\mu}$. \\
\deleted{The key point in these calculations is the necessity of canonical duals to quantize a system. In the second-class set, the canonical duals are either outside the subspace or inside. The quantization process happens on the duals inside the subspace.
But in the first-class set, there are canonical duals that one side of the canonical duality is inside the subspace and the other is outside the subspace. This problem disrupts the quantization process.
We have reduced the subspace to the extent that there is no defective canonical duality. Gauge symmetry enables this reduction, which is one of the characteristics of first-class constraints.}

\added{A central aspect of these considerations is the necessity of having complete canonical pairs for quantization. For second-class constraints, the canonical pairs are either entirely within or entirely outside the reduced subspace, and quantization proceeds over the pairs contained within the subspace. However, in the presence of first-class constraints, certain canonical pairs become defective—one member lies within the subspace while the other lies outside—disrupting the quantization process. To resolve this, the subspace is further reduced to eliminate such incomplete canonical structures. This reduction is facilitated by gauge symmetry, a fundamental feature associated with first-class constraints.}

\section{Applying external constraints}
\deleted{Let us consider these external constraints applied to the fields:}\added{To develop our new framework, we consider the application of external constraints to the fields in the following form:}
\be\label{3.2}
f_{\alpha}(q,\dot{q})=b_{\alpha i}(q)\dot{q}_{i}+c_{\alpha}(q),
\ee
\deleted{which the discrete spacial coordinates are included in the index $i$.}\added{where the index $i$ includes both discrete labels for the degrees of freedom and the spatial coordinates.} \deleted{For an original Lagrangian of the form}\added{We assume the original Lagrangian has the form}:
\be\label{3.1}
L=\frac{1}{2}A_{ij}(q)\dot{q}_{i}\dot{q}_{j}+B_{i}(q)\dot{q}_{i}-V(q),
\ee
where $\det{A_{ij}}\neq0$, we will incorporate the constraints as follows:
\be\label{3.3}
\tilde{L}=\frac{1}{2}A_{ij}(q)\dot{q}_{i}\dot{q}_{j}+B_{i}(q)\dot{q}_{i}-V(q)-v_{\alpha}(b_{\alpha i}(q)\dot{q}_{i}+c_{\alpha}(q)),
\ee
such that the quantities $v_{\alpha}$ are considered as additional degrees of freedom within the deformed Lagrangian.\\
If we consider: 
\be\label{3.3a}
b_{\alpha i} \neq 0,
\ee
we find new canonical momenta: 
\be\label{3.4}
p_{i}=A_{ij}(q)\dot{q}_{j}+B_{i}(q)-v_{\alpha}b_{\alpha i}(q), 
\ee  
and the canonical momentum of  $v_{\alpha}$ is zero as a primary constraint:
\be\label{3.5}
w_{\alpha}=0.
\ee
And if we rewrite $f_{\alpha}$ based on $q$, $p$ we will have our secondary constraints:
\be\label{3.6}
\phi_{\alpha}(q,p,v)= f_{\alpha}(q,\dot{q}(q,p))=b_{\alpha i}(q)A^{-1}_{ij}(q)(p_{j}-B_{j}(q)+v_{\beta}b_{\beta j}(q))+c_{\alpha}(q)=0. 
\ee
Also, we can find a Hamiltonian with these canonical coordinates:
\be\label{3.7}
\tilde{H}=\dot{v}_{\alpha}w_{\alpha}+\frac{1}{2}A^{-1}_{ij}(q)(p_{i}-B_{i}(q)+v_{\alpha}b_{\alpha i}(q))(p_{j}-B_{j}(q)+v_{\beta}b_{\beta j}(q))+v_{\alpha}c_{\alpha }(q)+V(q). 
\ee 
Compared to equation \eqref{2.33} for the second-class constraints , the functional integral can be defined as:
\be\label{3.8}
\int\mathcal{D}q\mathcal{D}v\mathcal{D}p\mathcal{D}w \det \deleted{\cancel{[}} \added{\{} w_{\beta},\phi_{\alpha} \deleted{\cancel{]}} \added{\}}\delta(w_{\alpha})\delta(\phi_{\alpha}(q,p,v))e^{ i\int dt \;\dot{v}_{\alpha}w_{\alpha}+\dot{q}_{i}p_{i}-\tilde{H}(q,p,\dot{v},w)}.
\ee
We can derive $\deleted{\det}\deleted{\cancel{[}}\added{\{}w_{\beta},\phi_{\alpha}\deleted{\cancel{]}}\added{\}}=\frac{\partial\phi_{\alpha}}{\partial v_{\beta}}=b_{\alpha i}(q)A^{-1}_{ij}b_{\beta j}(q)$, which we call $E_{\alpha\beta}(q)$.
In continue we change $p_{i}$ to $p_{i}+B_{i}(q)-v_{\alpha}b_{\alpha i}(q) $ and the functional integral will be written as follows:
\bea\label{3.9}
\int\mathcal{D}q\mathcal{D}v\mathcal{D}p\det( E_{\alpha\beta})\delta(b_{\alpha i}A^{-1}_{ij}p_{j}+c_{\alpha})~~~~~~~~~~~~~~~~~~~~~~~~~~~~~~~~~~~~~~ \nonumber\\
exp\bigg[ i\int dt \;\dot{q}_{i}(p_{i}+B_{i}(q)-v_{\alpha}b_{\alpha i}(q))-\frac{1}{2}A^{-1}_{ij}(q)p_{i}p_{j}-v_{\alpha}c_{\alpha }(q)-V(q)\bigg].
\eea
If we replace the delta functions with the Fourier integrals $\int\mathcal{D}\lambda e^{i\int dt\;\lambda_{\alpha}(b_{\alpha i}A^{-1}_{ij}p_{j}+c_{\alpha})}$ ($\lambda_{\alpha}$ is the Lagrange multiplier for the secondary constraint.)
and changing $v_{\alpha}$ to $v_{\alpha}+\lambda_{\alpha}$, we will have:
\bea\label{3.10}
\int\mathcal{D}q \mathcal{D}v\det( E_{\alpha\beta}) e^{i\int dt\;B_{i}(q)\dot{q}_{i}-V(q)-v_{\alpha}b_{\alpha i}(q)\dot{q}_{i}-v_{\alpha}c_{\alpha }(q)}\mathcal{D}\lambda\mathcal{D}p~~~~~~~\nonumber\\
exp\bigg[i\int dt\;\lambda_{\alpha}b_{\alpha i}(q)A_{ij}^{-1}(q)p_{j}-\lambda_{\alpha}b_{\alpha i}(q)\dot{q}_{i}+\dot{q}_{i}p_{i}-\frac{1}{2}A^{-1}_{ij}(q)p_{i}p_{j}\bigg].
\eea
By integrating over $p$ we have:
\bea\label{3.11}
\int\mathcal{D}q \mathcal{D}v\det( E_{\alpha\beta})\sqrt{\det(A_{ij})} e^{i\int dt\;B_{i}(q)\dot{q}_{i}-V(q)-v_{\alpha}b_{\alpha i}(q)\dot{q}_{i}-v_{\alpha}c_{\alpha }(q)}\mathcal{D}\lambda \nonumber\\
exp\bigg[i\int dt\;-\lambda_{\alpha}b_{\alpha i}(q)\dot{q}_{i}+\frac{1}{2}A_{ij}^{-1}[\lambda_{\alpha}b_{\alpha i}(q)+\dot{q}_{k}A_{ik}(q)][\lambda_{\alpha}b_{\alpha j}(q)+\dot{q}_{l}A_{jl}(q)]\bigg],
\eea
and also integrating over $\lambda$:
\be\label{3.12}
\int\mathcal{D}q \mathcal{D}v\sqrt{\det( E_{\alpha\beta})}\sqrt{\det(A_{ij})} e^{i\int dt\;L(q,\dot{q})-v_{\alpha}f_{\alpha}(q,\dot{q})}.
\ee
Finally by integrating over $v$, we will obtain this generating functional:
\be\label{3.13}
Z(J)=\int\mathcal{D}q\sqrt{\det( E_{\alpha\beta})}\sqrt{\det(A_{ij})}\delta(f(q,\dot{q})) e^{i\int dt\;L(q,\dot{q})+J_{i}q_{i}}.
\ee
\deleted{If you concentrate on the delta functionals in the above equation, you will observe that the number of degrees of freedom is reduced by the number of applied constraints.}\added{By examining the delta functionals in the above expression, it becomes evident that the external constraint enters the path integral in its original classical form, without modification. This implies that the constraint is not subjected to quantum fluctuations and remains enforced exactly at the quantum level, just as it is classically.} \deleted{However, from a computational perspective, the presence of delta functionals can sometimes complicate problem-solving.}

\deleted{To find a functional integral without delta functionals, we can return to the equation\eqref{3.8}.}
\added{Naturally, one can reformulate the partition function such that the delta functional no longer appears explicitly, thereby internalizing the constraints and embedding them intrinsically into the quantum description. Restarting from equation \eqref{3.8}, we eliminate the factor}
\deleted{We can eliminate} $\mathcal{D}v\det(\frac{\partial\phi_{\alpha}}{\partial v_{\beta}})\delta(\phi_{\alpha})$ and \deleted{also} replace $v_{\alpha}$ with its \deleted{value}\added{solution} obtained
from \deleted{the} equation\eqref{3.6}. By changing $p_{i}$ to $p_{i}+B_{i}$ we \deleted{will have}\added{arrive at}:
\bea\label{3.14}
\int\mathcal{D}q\; e^{i\int dt\;-V(q)+\dot{q}_{i}B_{i} }\mathcal{D}p~~~~~~~~~~~~~~~~~~~~~~~~~~~~~~~~~~~~~~~~~~~~~~~~~~~~\nonumber\\
exp\bigg[ i\int dt \;\dot{q}_{i}p_{i}-\frac{1}{2}A^{-1}_{ij}(q)[p_{i}+\deleted{\cancel{v}}\added{u}_{\beta}b_{\beta i}(q)][p_{j}+\deleted{\cancel{v}}\added{u}_{\alpha}b_{\alpha j}(q)]-\deleted{\cancel{v}}\added{u}_{\alpha}c_{\alpha }(q)\bigg],
\eea
\deleted{such that}\added{where}, $\deleted{\cancel{v}}\added{u}_{\alpha}=-E^{-1}_{\alpha\beta}c_{\beta}-E^{-1}_{\alpha\beta}b_{\beta i}A^{-1}_{ij}p_{j}$. \added{By substituting this expression for} $u_{\alpha}$ into \eqref{3.14}, \added{the} partition function \deleted{will be as the following form}\added{takes the form:} \deleted{And if $A_{ij}$ and $b_{\alpha i}$ are constant parameters, we can have an ordinary functional path integral.} 
\added{
\bea\label{3.14a}
\int\mathcal{D}q\; exp\bigg(i\int dt\;-(V(q)+U(q))+\dot{q}_{i}(B_{i}(q)+H_{i}(q))\bigg) \mathcal{D}p\nonumber\\ ~exp\bigg[ i\int dt \;\dot{q}_{i}p_{i}-\frac{1}{2}F_{ij}(q)p_{i}p_{j}\bigg],
\eea}
\added{where the functions $U(q),H_{i}(q),$ and$F_{ij}(q)$ are derived from the original data:\\
$V(q),B_{i}(q),A_{ij}(q),c_{\alpha}(q),$ and $b_{\alpha i}(q)$. This final representation shows that the external constraints have been fully internalized within the effective action. The path integral in \eqref{3.14a} encodes the constrained dynamics through modified scalar potentials and redefined kinetic coefficients, thereby eliminating the need for explicit delta functionals or auxiliary integration variables.}

Next, consider equation \eqref{3.2} under the following condition:
\be\label{3.15}
b_{\alpha i}=0.
\ee 
\deleted{This case has been previously reviewed}\added{A similar case has previously been examined in}\cite{taro,toms}; however, we will analyze it using the proposed method
.\\
The canonical momenta are:
\be\label{3.16}
p_{i}=A_{ij}(q)\dot{q}_{j}+B_{i}(q), 
\ee
and $w_{\alpha}=0$ which are considered primary constraints. The secondary constraints are $c_{\alpha}(q)=0$. Since the time derivative of these secondary constraints are constraints $\dot{c}_{\alpha}(q)=\dot{q}_{i}\frac{\delta c_{\alpha}(q)}{\delta q_{i}}$,
we also have tertiary constraints:
\be\label{3.17}
d_{\alpha}(q,p)=A^{-1}_{ij}(q)(p_{i}-B_{i}(q))\frac{\delta c_{\alpha}(q)}{\delta q_{j}}=0.
\ee
The Poisson brackets between the primary constraints and other constraints indicate that this set of constraints is first-class:
\be\label{3.18}
\{w_{\alpha}, c_{\beta}\}=\{w_{\alpha}, d_{\beta}\}=0.
\ee
However, according to the recent equations, we are only satisfied with the canonical transformation of $(q,p)$ and keep the pair $(v,w)$ without transformation. 
Because of the primary constraints, $v_{\alpha}$ is not a quantum coordinate. As a result, with the help of equations\eqref{2.24} and \eqref{2.31}, we can write the new functional integral as follows: 
\be\label{3.19}
\int\mathcal{D}q\mathcal{D}p \det\{d_{\beta}(q),c_{\alpha}(q,p)\}\delta(c_{\alpha}(q))\delta(d_{\beta}(q,p))e^{ i\int dt \;\dot{q}_{i}p_{i}-\tilde{H}(q,p)},
\ee
such that,
\be\label{3.20}
\tilde{H}=\frac{1}{2}A^{-1}_{ij}(q)(p_{i}-B_{i}(q))(p_{j}-B_{j}(q))+v_{\alpha}c_{\alpha }(q)+V(q). 
\ee 
The presence of $\delta(c_{\alpha})$ removes expression $v_{\alpha}c_{\alpha }$ in the Hamiltonian and we will have:
\be\label{3.21}
\int\mathcal{D}q\mathcal{D}p \det\{d_{\beta},c_{\alpha}\}\delta(c_{\alpha}(q))\delta(d_{\beta}(q,p))e^{ i\int dt \;\dot{q}_{i}p_{i}-\frac{1}{2}A^{-1}_{ij}(q)(p_{i}-B_{i}(q))(p_{j}-B_{j}(q))-V(q)},
\ee
so that,
\be\label{3.22}
\{d_{\beta},c_{\alpha}\}=\{A^{-1}_{ij}p_{i}\frac{\delta c_{\beta}}{\delta q_{j}},c_{\alpha}\}=A^{-1}_{ij}\frac{\delta c_{\alpha}}{\delta q_{i}}\frac{\delta c_{\beta}}{\delta q_{j}}=D_{\alpha\beta}.
\ee
By changing $p_{i}$ to $p_{i}+B_{i}$, equation \eqref{3.21} becomes as follows:
\be\label{3.23}
\int\mathcal{D}q\mathcal{D}p \det(D_{\alpha\beta})\delta(c_{\alpha})\delta(A^{-1}_{ij}p_{i}\frac{\delta c_{\alpha}}{\delta q_{j}})e^{ i\int dt \;\dot{q}_{i}(p_{i}+B_{i})-\frac{1}{2}A^{-1}_{ij}p_{i}p_{j}-V}.
\ee
We write the Dirac delta functional of the tertiary constraint based on the integral over the Lagrange multiplier. By integrating over the canonical momentums, we get:
\be\label{3.24}
\int\mathcal{D}q \det(D_{\alpha\beta})\sqrt{\det(A_{ij})}\delta(c_{\alpha})e^{ i\int dt \;\dot{q}_{i}B_{i}-V}\int\mathcal{D}\lambda e^{\frac{i}{2}\int dt\, A_{ij}(\dot{q}_{i}+\lambda_{\alpha}A^{-1}_{ik}\frac{\delta c_{\alpha}}{\delta q_{k}})(\dot{q}_{j}+\lambda_{\beta}A^{-1}_{jl}\frac{\delta c_{\beta}}{\delta q_{l}})}.
\ee
After integrating over $\lambda$, we derive the generating functional as follows:
\be\label{3.25}
Z(J)=\int\mathcal{D}q \sqrt{\det(D_{\alpha\beta})}\sqrt{\det(A_{ij})}\delta(c_{\alpha})e^{ i\int dt\, L+J_{i}q_{i}-i\int dt \;\frac{1}{2}\dot{q}_{i}G_{ij}\dot{q}_{j}},
\ee
so that:
\be\label{3.26}
G_{ij}=D^{-1}_{\alpha\beta}\frac{\delta c_{\alpha}}{\delta q_{i}}\frac{\delta c_{\beta}}{\delta q_{j}}.
\ee
Comparing equations \eqref{3.25} and \eqref{3.13} highlights a significant advantage where $b_{\alpha i}\neq0$. In such cases, the kinetic part of the Lagrangian remains unaffected, allowing all degrees of freedom to retain their dynamical character.

\section{Spin-one state as an external constraint}
In this section, we \deleted{introduce a clear}\added{present a concrete} example \deleted{for the}\added{of an} external constraint. Lorentz vectors simultaneously \deleted{represent}\added{describe both} the singlet state of spin-zero and the triplet state of spin-one. We can \deleted{define}\added{isolate} a \deleted{state with} spin-one \added{state }\deleted{only} and impose this state \deleted{through}\added{as} an external constraint.
\deleted{For}\added{To achieve} this \deleted{purpose}, we \deleted{must impose a constraint requiring the vector field to be orthogonal}\added{enforce a condition that requires the vector field to be orthogonal} to the momentum direction\deleted{, which}\added{.This constraint} can be expressed \deleted{as follows} in momentum coordinates \added{as}:
\be\label{3.27}
p_{\mu}\tilde\chi^{\mu}(p)=0.
\ee
In \deleted{other words}\added{this formulation}, the vector field is decomposed into components \added{both}  perpendicular and parallel to the momentum direction: $\tilde\chi_{\perp}(p)$ and $\tilde\chi_{\parallel}(p)$\added{, respectively.} \deleted{So that}\added{The component} $\tilde\chi_{\perp}(p)$ is treated as a quantized field, while $\tilde\chi_{\parallel}(p)$ is regarded as a classical field constrained to \deleted{have a numerical value of}\added{be} zero. 
\deleted{And} In the time-space coordinate, \deleted{our}\added{the} constraint \deleted{will be}\added{is then expressed as}:
\be\label{3.28}
\partial_{\mu}\chi^{\mu}(x)=0.
\ee
One example of a field that describes particles with spin-one is the Yang-Mills field with broken gauge symmetry, \added{or equivalently the Proca field}, where the \deleted{specified}\added{ associated} constraint naturally \deleted{emerges}\added{arises} from its equations of motion. \deleted{However, it represents intermediate particles that}\added{This field typically represents intermediate particles that} facilitate the interaction between matter particles.
However, \deleted{if we want}\added{to describe} matter particles with spin-one, we \added{suggest} \deleted{must} impos\deleted{e}\added{ing} this constraint externally on \deleted{the} Lorentzian vector fields \added{governed by a Klein-Gordon-type action.}

We can add the constraint introduced earlier to any Lagrangian that involves a Lorentzian vector field. For simplicity, we now consider the following Lagrangian density:
\be\label{3.29}
\mathcal{L}=\frac{1}{2}\partial_{\nu}\chi^{\mu}\partial^{\nu}\chi_{\mu}-\frac{1}{2}m^{2}\chi^{\mu}\chi_{\mu}+\mathcal{L}_{int}.
\ee
By comparison with equation \eqref{3.1}, we express the Lagrangian as follows:
\be\label{3.30} 
L=\int d^3x d^3x' \frac{1}{2}\delta^3(\vec{x}-\vec{x'})\eta_{\mu\nu}\dot{\chi}^{\mu}(t,\vec{x})\dot{\chi}^{\nu}(t,\vec{x'})+...
\ee
And if we rewrite the equation \eqref{3.28} in the form of the equation \eqref{3.2}, we will have:
\be\label{3.31}
\int d^3\vec{x'} \delta^3(\vec{x}-\vec{x'}) \eta^0_{\mu}\dot{\chi}^{\mu}(t,\vec{x'})+\vec{\nabla}.\vec{\chi}(t,\vec{x})=0.
\ee
Thus, we can obtain:
\be\label{3.32}
E(\vec{x},\vec{x'})=\int d^3yd^3z  \delta^3(\vec{x}-\vec{y})\eta_{0\mu} \delta^3(\vec{x}-\vec{x'}) \delta^3(\vec{y}-\vec{z})\eta^{\mu\nu} \delta^3(\vec{z}-\vec{x'})\eta_{0\nu}= \delta^3(\vec{x}-\vec{x'}).
\ee
As a result, the equation \eqref{3.13} becomes:
\be
Z_{int}(J)=\int\mathcal{D}\chi^{\mu}\delta(\partial_{\mu}\chi^{\mu}(x))e^{i\int d^4x \frac{1}{2}\partial_{\nu}\chi^{\mu}(x)\partial^{\nu}\chi_{\mu}(x)-\frac{1}{2}m^2\chi^{\mu}(x)\chi_{\nu}(x)+\mathcal{L}_{int}+J_{\mu}(x)\chi^{\mu}(x)}.
\ee
In the perturbation method, the interaction term is extracted from the exponential function and converted into coefficients that are functionals of the fields. In this case, we only need to calculate the n-point functions on the interaction-free background, and we will use the generating functional of the free Lagrangian density:
\be\label{3.33}
Z(J)=\int\mathcal{D}\chi^{\mu}\delta(\partial_{\mu}\chi^{\mu}(x))e^{i\int d^4x \frac{1}{2}\partial_{\nu}\chi^{\mu}(x)\partial^{\nu}\chi_{\mu}(x)-\frac{1}{2}m^2\chi^{\mu}(x)\chi_{\nu}(x)+J_{\mu}(x)\chi^{\mu}(x)}.
\ee
To obtain this generating functional we use the momentum Fourier transformation:
\be\label{3.33}
Z(J)=\int\mathcal{D}\tilde\chi^{\mu}\delta(p_{\mu}\tilde\chi^{\mu}(p))e^{i\int d^4p \frac{1}{2}\tilde\chi^{\mu}(p)(p^2-m^2)\tilde\chi_{\mu}(p)+\tilde J_{\mu}(p)\tilde\chi^{\mu}(p)}.
\ee
To incorporate the Dirac delta function into the path integral, we utilize the following equation:
\be\label{3.34}
\tilde\chi^0=-\frac{p_{i}}{p_{0}}\tilde\chi^i,
\ee
which is obtained from the equation \eqref{3.27}. In this case, the generating functional will be written as follows:
\be\label{3.35}
\int\mathcal{D}\tilde\chi^{i}exp\bigg(i\int d^4p \frac{1}{2}\tilde\chi^{i}(p)(p^2-m^2)(\eta_{ij}+\frac{p_{i}p_{j}}{p_{0}^2})\tilde\chi_{j}(p)+(\tilde J_{i}(p)-\frac{p_{i}}{p_{0}}\tilde J_{0}(p))\tilde\chi^{i}(p)\bigg).
\ee
If we transform $\tilde\chi$ like this:
\be\label{3.36}
\tilde\chi^i \rightarrow\tilde\chi^i-\frac{1}{p^2-m^2}(\eta_{ij}+\frac{p_{i}p_{j}}{p_{0}^2})^{-1}(\tilde J_{j}-\frac{p_{j}}{p_{0}}\tilde J_{0}),
\ee
and considering that:
\be\label{3.37}
(\eta_{ij}+\frac{p_{i}p_{j}}{p_{0}^2})^{-1}=\eta^{ij}-\frac{p^{i}p^{j}}{p^2},
\ee
the generating functional will be:
\be\label{3.38}
\int\mathcal{D}\tilde\chi^{i}e^{i\int d^4p \frac{1}{2}\tilde\chi^{i}(p)(p^2-m^2)(\eta_{ij}+\frac{p_{i}p_{j}}{p_{0}^2})\tilde\chi_{j}(p)} exp\bigg(\frac{-i}{2}\int d^4p(\tilde J_{i}-\frac{p_{i}}{p_{0}}\tilde J_{0})\frac{(\eta^{ij}-\frac{p^{i}p^{j}}{p^2})}{p^2-m^2}(\tilde J_{j}-\frac{p_{j}}{p_{0}}\tilde J_{0})\bigg).
\ee
Next, the generating functional is written as follows:
\be\label{3.39}
Z(J)\propto exp\bigg(-\frac{i}{2}\int d^4p\tilde J_{\mu}(p)\frac{\eta^{\mu\nu}-p^{\mu}p^{\nu}/p^2}{p^2-m^2}\tilde J_{\nu}(p)\bigg).
\ee
According to the obtained result, the two-point function in momentum coordinates is expressed as follows:
\be\label{3.40}
G^{\mu\nu}(p)=\frac{i}{p^2-m^2}(\eta^{\mu\nu}-\frac{p^{\mu}p^{\nu}}{p^2}).
\ee
\added{This feature resembles the structure observed in the Landau gauge propagator, where the propagator in momentum space includes a projection onto the transverse components of the momentum. This projection ensures that only the physical degrees of freedom are included, excluding the unphysical longitudinal modes. }\deleted{The key point of this propagator function is that its determinant is zero, making it non-invertible. As a result, this function cannot be considered as the Green's function derived from any linear equation of motion, which makes it non-trivial.}\added{A key aspect of these similar propagators is that their determinant vanishes, rendering them non-invertible. This highlights the non-trivial nature of the propagator, as it cannot be regarded as the Green's function derived from any linear equation of motion. }

\section{Applying external constraints on the system with a singular action}
We can consider a Lagrangian with internal constraints: 
\be\label{4.1}
L=\frac{1}{2}A_{ij}(q,Q)\dot{q}_i\dot{q}_j+B_{i}(q,Q)\dot{q}_i+B_{a}(q,Q)\dot{Q}_a-V(q,Q).
\ee
Furthermore, if we imply the constraints \eqref{3.2} to this Lagrangian, we obtain:
\bea\label{4.2}
\tilde{L}=L-v_{\alpha}f_{\alpha}=~~~~~~~~~~~~~~~~~~~~~~~~~~~~~~~~~~~~~~~~~~~~~~~~~~~~~~~~~~~~~~~~~~~~~~~~~~~~~~~~\nonumber\\
\frac{1}{2}A_{ij}(q,Q)\dot{q}_{i}\dot{q}_{j}+B_{i}(q,Q)\dot{q}_{i}+B_{a}(q,Q)\dot{Q}_a-V(q,Q)~~~~~~~~~~~~~~~~~~~~~~~~~~~~~~~\nonumber\\
-v_{\alpha}(b_{\alpha i}(q,Q)\dot{q}_{i}+b_{\alpha a}(q,Q)\dot{Q}_{a}+c_{\alpha}(q,Q)).~~~~~~~~~~~~~~~~~~~~
\eea
Therefore the canonical momenta are:
\bea\label{4.3}
p_{i}(q,Q,v,\dot{q})=A_{ij}(q,Q)\dot{q}_{j}+B_{i}(q,Q)-v_{\alpha}b_{\alpha i}(q,Q),\nonumber\\
P_{a}=B_{a}(q,Q)-v_{\alpha}b_{\alpha a}(q,Q),~W_{\alpha}=0.~~~~~~~~~~~~~~~~~~
\eea
The equations of motion of $v_{\alpha}$ are:
\be\label{4.3a}
f_{\alpha}(q,Q,\dot{q}(q,p,Q,v),\dot{Q}_{a})=0.
\ee
For any $\dot{Q}_{a}$ obtained from these equations, one constraint is lost.
If we set $b_{\alpha a}=0$, all the equations for $v_{\alpha}$ can be treated as secondary constraints.
\be\label{4.4}
\phi_{\alpha}(q,p,Q,v)=f_{\alpha}(q,Q,\dot{q}(q,p,Q,v)),
\ee
such that using \eqref{4.3}, the time derivatives of the degrees of freedom are:
\be\label{4.5}
\dot{q}_{i}(q,p,Q,v)=A^{-1}_{ij}(q,Q)[p_{j}-B_{j}(q,Q)+v_{\alpha}b_{\alpha j}(q,Q)].
\ee
And in this case, the equations of motion of $Q_{a}$ are:
\be\label{4.6}
g_{a}(q,Q,v,\dot{q})=\frac{\partial\tilde{L}(q,Q,v,\dot{q})}{\partial Q_{a}}-\frac{\partial B_{a}(q,Q)}{\partial Q_{b}}\dot{Q}_{b}-\frac{\partial B_{a}(q,Q)}{\partial q_{i}}\dot{q}_{i}=0.
\ee
In these equations, the expressions containing $\dot{Q}_{a}$ are as follows:
\be\label{4.7}
(\frac{\partial B_{a}}{\partial Q_{b}}-\frac{\partial B_{b}}{\partial Q_{a}})\dot{Q}_{a}.
\ee
If $\det(\frac{\partial B_{a}}{\partial Q_{b}}-\frac{\partial B_{b}}{\partial Q_{a}})=0$, the set of $\dot{Q}_{a}$ remains undetermined from the solution of the equations \eqref{4.6}. With a variable transformation on $Q_{a}$\added{to $Q_{A}$ and $Q_{N}$                                           such that $\frac{\partial\tilde{L}}{\partial \dot{Q}_{A}}=0$, while $\dot{Q}_{N}$ are determined.} 
we represent the undetermined members of this set as \added{$\dot{Q}_{A}$} \deleted{$\dot{Q}_{A}=\frac{\partial Q_{A}}{\partial Q_{a}}\dot{Q_{a}}$}. Therefore, we will have new secondary constraint equations  corresponding to the number of these undetermined variables.
We represent these new secondary constraint equations as follows:
\be\label{4.8}
\chi_{A}(q,p,Q,v)=g_{A}(q,Q,v,\dot{q}(q,p,Q,v))=\frac{\partial\tilde{L}(q,Q,v,\dot{q})}{\partial Q_{A}}=0.
\ee
Similar to equation \eqref{3.7} the hamiltonian is: 
\bea\label{4.9a}
\tilde{H}(q,p,Q,v)=\frac{1}{2}A^{-1}_{ij}(q,Q)(p_{i}-B_{i}(q,Q)+v_{\alpha}b_{\alpha i}(q,Q))(p_{j}-B_{j}(q,Q)+v_{\beta}b_{\beta j}(q,Q))\nonumber\\ 
+v_{\alpha}c_{\alpha }(q,Q)+V(q,Q).~~~~~~~~~~~~~~~~~~~~~~~~~~~~~~~~~~~~~~~~~~~~~~~~~~~~~~~~~~~
\eea
Using equation \eqref{2.43}, we will have:

\bea\label{4.10a}
\int \mathcal{D}q_{i}\mathcal{D}p_{i}\mathcal{D}Q_{a}\mathcal{D}v\det\{\Psi_{\mu},X_{\nu}\}(\det\{\Gamma_{M},\Gamma_{N}\})^{\frac{1}{2}}\delta(f_{\alpha}(q,Q,\dot{q}(q,p,Q,v)))\delta(g_{A}(q,Q,v,\dot{q}(q,p,Q,v)))\nonumber\\
exp\bigg(i\int dt \dot{q}_{i}p_{i}+\dot{Q}_{a}B_{a}-\tilde{H}(q,p,Q,v)\bigg).~~~~~~~~~~~~~~~~~~~~~~~~~~~~~~~~~
\eea
Changing $p_{i}$ to $p_{i}+B_{i}(q,Q)-v_{\alpha}b_{\alpha i}(q,Q)$ and using equation \eqref{4.5}, we have:
\bea\label{4.11a}
\int \mathcal{D}q_{i}\mathcal{D}p_{i}\mathcal{D}Q_{a}\mathcal{D}v\det\{\Psi_{\mu},X_{\nu}\}(\det\{\Gamma_{M},\Gamma_{N}\})^{\frac{1}{2}}\delta(f_{\alpha}(q,Q,A^{-1}_{ij}(q,Q)p_{j}))\delta(g_{A}(q,Q,v,A^{-1}_{ij}(q,Q)p_{j}))\nonumber\\
exp\bigg(i\int dt \dot{q}_{i}(p_{i}+B_{i}(q,Q)-v_{\alpha}b_{\alpha i}(q,Q))+\dot{Q}_{a}B_{a}-\frac{1}{2}A^{-1}_{ij}(q,Q)p_{i}p_{j}-v_{\alpha}c_{\alpha }(q,Q)-V(q,Q)\bigg).
\eea
With the last variable change made in the set of momenta, we obtain:
\be\label{4.12a}
\frac{\partial \phi_{\alpha}}{\partial v_{\beta}}=0.
\ee
However, to prove the following equation: 
\be\label{4.12b}
\frac{\partial \chi_{A}}{\partial v_{\beta}}=0,
\ee
we need sufficient conditions. The most general sufficient conditions, independent of the form of the coefficients in the Lagrangian defined in \eqref{4.1}, are as follows:
\be\label{4.13a}
\frac{\partial b_{\alpha i}}{\partial Q_{A}}=\frac{\partial c_{\alpha}}{\partial Q_{A}}=0,~~~~~b_{\alpha a}=0.
\ee
Therefore, the measure of the path integral and the Dirac delta functionals are not dependent on $v_{\alpha}$, allowing us to express the functional integral as follows:
 \bea\label{4.15a}
\int \mathcal{D}q_{i}\mathcal{D}p_{i}\mathcal{D}Q_{a}\det\{\Psi_{\mu},X_{\nu}\}(\det\{\Gamma_{M},\Gamma_{N}\})^{\frac{1}{2}}\nonumber\\
\delta(f_{\alpha}(q,Q,A^{-1}_{ij}(q,Q)p_{j}))\delta(g_{A}(q,Q,A^{-1}_{ij}(q,Q)p_{j}))\delta(f_{\alpha}(q,Q,\dot{q}))\nonumber\\
exp\bigg(i\int dt \dot{q}_{i}(p_{i}+B_{i}(q,Q))+\dot{Q}_{a}B_{a}-\frac{1}{2}A^{-1}_{ij}(q,Q)p_{i}p_{j}-V(q,Q)\bigg),
\eea
where the term $ \delta(f_{\alpha}(q,Q,\dot{q}))$ can be found similarly to equation\eqref{3.13}.

\section{A choice of Lagrangian coefficients}

\deleted{If the form of the Lagrangian coefficients is more specific, the sufficient conditions for finding the $ \delta(f_{\alpha}(q,\dot{q}))$ term will offer more possibilities. For example, if all $B_{a}$ are zero, the set of constraints, like in \eqref{4.8}, is:}\added{When the Lagrangian coefficients are chosen with a more specific structure, the sufficient conditions for determining the $\delta(f_{\alpha}(q,\dot{q}))$ term allow for greater flexibility. For instance, if all $B_{a}$ vanish, the resulting set of constraints, as in equation \eqref{4.8}, becomes:}

\be\label{5.1}
\chi_{a}=g_{a}(q,Q,v,\dot{q}(q,Q,v,p)),
\ee
where
\be\label{5.2}
g_{a}(q,Q,v,\dot{q})=\frac{\partial\tilde{L}(q,Q,v,\dot{q})}{\partial Q_{a}}.
\ee
The primary constraints are:
\be\label{5.3}
P_{a}=0, w_{\alpha}=0.
\ee
Thus, using equation \eqref{2.33}, we have:
\be\label{4.9}
\int\mathcal{D}q\mathcal{D}Q\mathcal{D}v\mathcal{D}p\det(E)\delta(\varphi_{\alpha}(q,p,Q,v))\delta(\chi_{\alpha}(q,p,Q,v))e^{ i\int dt \dot{q}_{i}p_{i}-\tilde{H}(q,p,Q,v)},
\ee
where,
\be\label{4.10}
E_{\alpha \beta}=\frac{\partial\varphi_{\alpha}}{\partial v_{\beta}}=\frac{\partial f_{\alpha}}{\partial \dot{q}_{i}}\frac{\partial \dot{q}_{i}}{\partial v_{\beta}}=b_{\alpha i}A^{-1}_{ij}b_{\beta j},
\ee
\be
E_{\alpha a}=\frac{\partial\varphi_{\alpha}}{\partial Q_{a}}=\frac{\partial f_{\alpha}}{\partial Q_{a}}+\frac{\partial f_{\alpha}}{\partial \dot{q}_{i}}\frac{\partial \dot{q}_{i}}{\partial Q_{a}}\nonumber\\
\ee

\be\label{4.11}
=\frac{\partial f_{\alpha}}{\partial Q_{a}}+b_{\alpha i}\frac{\partial A^{-1}_{ij}}{\partial Q_{a}}(p_{j}-B_{j}+v_{\alpha}b_{\alpha j})-b_{\alpha i}A^{-1}_{ij}(\frac{\partial B_{j}}{\partial Q_{a}}-v_{\alpha}\frac{\partial b_{\alpha j}}{\partial Q_{a}}),
\ee
and,
\be
E_{a\alpha}=\frac{\partial\chi_{a}}{\partial v_{\alpha}}=\frac{\partial g_{a}}{\partial v_{\alpha}}+\frac{\partial g_{a}}{\partial \dot{q}_{i}}\frac{\partial \dot{q}_{i}}{\partial v_{\alpha}}\nonumber\\
\ee
\be
=-\frac{\partial f_{\alpha}}{\partial Q_{a}}+\frac{\partial^2\tilde{L}}{\partial Q_{a}\partial\dot{q}_{i}}A^{-1}_{ij}b_{\alpha j}\nonumber\\
\ee
\be\label{4.13}
=-\frac{\partial f_{\alpha}}{\partial Q_{a}}-b_{\alpha i}\frac{\partial A^{-1}_{ij}}{\partial Q_{a}}(p_{j}-B_{j}+v_{\alpha}b_{\alpha j})+b_{\alpha i}A^{-1}_{ij}(\frac{\partial B_{j}}{\partial Q_{a}}-v_{\alpha}\frac{\partial b_{\alpha j}}{\partial Q_{a}}),
\ee
which $E_{a\alpha}=-E_{\alpha a}$,
and likewise,
\be
E_{ab}=\frac{\partial\chi_{a}}{\partial Q_{b}}=\frac{\partial g_{a}}{\partial Q_{b}}+\frac{\partial g_{a}}{\partial \dot{q}_{i}}\frac{\partial \dot{q}_{i}}{\partial Q_{b}}\nonumber\\
\ee
\bea\label{4.14}
=\frac{\partial g_{a}}{\partial Q_{b}}-A^{-1}_{ij}\bigg(\frac{\partial B_{i}}{\partial Q_{a}}-v_{\alpha}\frac{\partial b_{\alpha i}}{\partial Q_{a}}+A^{-1}_{lk}\frac{\partial A_{il}}{\partial Q_{a}}(p_{k}-B_{k}+v_{\alpha}b_{\alpha k})\bigg)\nonumber\\
\bigg(\frac{\partial B_{j}}{\partial Q_{a}}-v_{\alpha}\frac{\partial b_{\alpha j}}{\partial Q_{a}}+A^{-1}_{lk}\frac{\partial A_{jl}}{\partial Q_{a}}(p_{k}-B_{k}+v_{\alpha}b_{\alpha k})\bigg).
\eea
In the equations \eqref{4.9}, \eqref{4.13} and \eqref{4.14}, if we replace $p_{k}-B_{k}+v_{\alpha}b_{\alpha k}$ with $p_{k}$, we obtain the following result:
\be\label{4.15}
\int\mathcal{D}q\mathcal{D}Q\mathcal{D}v\mathcal{D}p\det(E)\delta(\varphi_{\alpha})\delta(\chi_{a})e^{ i\int dt  \dot{q}_{i}(p_{i}+B_{i}-v_{\alpha}b_{\alpha i})-\frac{1}{2}A^{-1}_{ij}(q,Q)p_{i}p_{j}-v_{\alpha}c_{\alpha }(q,Q)-V(q,Q)}.
\ee
If we focus on equations \eqref{4.13} and \eqref{4.14}, we can observe that the term $v_{\alpha}\frac{\partial b_{\alpha j}}{\partial Q_{a}}$  is a functional of $v_{\alpha}$. Additionally, the term $v_{\alpha}\frac{\partial^2c_{\alpha}}{\partial Q_{a}\partial Q_{b}}$
appears in $\frac{\partial g_{a}}{\partial Q_{b}}$. In this context, the matrix $E$ will not be a functional of $v_{\alpha}$ if the following conditions hold:
\be\label{4.16}
\frac{\partial b_{\alpha j}}{\partial Q_{a}}=0,\frac{\partial^2c_{\alpha}}{\partial Q_{a}\partial Q_{b}}=0.
\ee
Therefore,
\bea\label{4.17}
\int\mathcal{D}q\mathcal{D}Q\mathcal{D}v\mathcal{D}p e^{-i\int dt v_{\alpha}(b_{\alpha i}(q)\dot{q}_{i}+c_{\alpha}(q,Q))}\det(E(q,Q,p))\delta(\varphi_{\alpha}(q,Q,p))\nonumber\\
\delta(\chi_{a}(q,Q,v=0)-v_{\alpha}\frac{\partial c_{\alpha}}{\partial Q_{a}})e^{ i\int dt  \dot{q}_{i}(p_{i}+B_{i}(q,Q))-\frac{1}{2}A^{-1}_{ij}(q,Q)p_{i}p_{j}-V(q,Q)}\nonumber\\
=\int\mathcal{D}\mu\mathcal{D}q\mathcal{D}Q\mathcal{D}v\mathcal{D}p e^{-i\int dt v_{\alpha}(b_{\alpha i}(q)\dot{q}_{i}+c_{\alpha}(q,Q))}\det(E(q,Q,p))\delta(\varphi_{\alpha}(q,Q,p))\nonumber\\
e^{i\int dt \mu_{a}\chi_{a}(q,Q,v=0)-\mu v_{\alpha}\frac{\partial c_{\alpha}}{\partial Q_{a}}}e^{ i\int dt  \dot{q}_{i}(p_{i}+B_{i}(q,Q))-\frac{1}{2}A^{-1}_{ij}(q,Q)p_{i}p_{j}-V(q,Q)}.
\eea
\deleted{Since $c_{\alpha}$ is a linear function of $Q_{a}$, we can replace $Q_{a}$ with $Q_{a}-\mu_{a}$ to eliminate $v_{\alpha}$ in the exponential function and therefore integrating over $v_{\alpha}$ to have the following functional integral:}\added{Since $c_{\alpha}$ depends linearly on $Q_{a}$ , we can perform a shift 
$Q_{a}\rightarrow Q_{a}-\mu_{a}$ to eliminate the dependence on  $v_{\alpha}$ in the exponential. This allows us to integrate over  $v_{\alpha}$ and obtain the following functional integral:}
\bea\label{4.18}
\int\mathcal{D}q\mathcal{D}Q\delta(f_{\alpha}(q,Q,\dot{q}))\int\mathcal{D}\mu\mathcal{D}p \det(E(q,Q-\mu,p))\delta(\varphi_{\alpha}(q,Q-\mu,p))\nonumber\\
exp \bigg[i\int dt \mu_{a}\chi_{a}(q,Q-\mu,p,0)+\dot{q}_{i}(p_{i}+B_{i}(q,Q-\mu))-\frac{1}{2}A^{-1}_{ij}(q,Q-\mu)p_{i}p_{j}-V(q,Q-\mu)\bigg],
\eea
in which, we can see $\delta(f_{\alpha}(q,Q,\dot{q}))$ similar to equation \eqref{3.13}.

\section{\added{Strength Field Derivation from External Constraints}}
\added{In this section, as an example, we consider a simple singular action with external constraints.}
In contrast to equation \eqref{4.13a}, a simple \deleted{example} \added{action} can be introduced by allowing the condition $b_{\alpha a}\neq0$\deleted{:\\}\added{.} A trivial singular Lagrangian density can be expressed in the following form:
\be\label{4.16}
\mathcal{L}[F]=-\frac{1}{4}F_{\mu\nu}(x)F^{\mu\nu}(x)+\mathcal{L}_{m},
\ee
where $F_{\mu\nu}$ serves as the fundamental anti symmetric degrees of freedom, and the equation of motion is trivially $F_{\mu\nu}(x)=0$. 
Consider the following external constraint:
\be\label{4.16a} 
\partial_{\mu}F^{\mu\nu}(x)=J^{\nu}(x).
\ee
This gives us:
\be\label{4.18a}
\tilde{\mathcal{L}}[F]=-\frac{1}{4}F_{\mu\nu}(x)F^{\mu\nu}(x)-v_{\mu}(x)(\partial_{\nu}F^{\nu\mu}(x)-J^{\mu}(x))+\mathcal{L}_{m}.
\ee
The primary constraints are:
\be\label{4.18ab}
W^{\mu}(x)=0, ~~~\Pi_{\mu<\nu}(x)\deleted{\cancel{-}}\added{+}\eta^0_{\mu}v_{\nu}(x)=0.
\ee
\added{The equations of motion for $v_{\mu}$ are:
\be\label{4.18ac}
\partial_{i}F^{0i}+J^{0}=0,
\ee
\be\label{4.18ad}
\partial_{0}F^{0i}+\partial_{i}F^{ij}-J^{j}=0,
\ee
and the equations of motion for $F^{\mu<\nu}$ are:
\be\label{4.18ae}
F_{i<j}-\partial_{i}v_{j}+\partial_{j}v_{i}=0,
\ee
\be\label{4.18af}
F_{0i}-\partial_{0}v_{i}+\partial_{i}v_{0}=0.
\ee
The latter equations present $F_{\mu\nu}$ as the strength field. Since we can not obtain the time derivatives of the degrees of freedom from the canonical momenta, the equations that involve time derivatives are not constraints. Therefore the secondary constraints will be:}
\be\label{4.18b}
\partial_{i}F^{0i}(x)+J^0(x)=0, ~~~ F^{i<j}(x)-\partial^{i}v^{j}(x)+\partial^{j}v^{i}(x)=0.
\ee
\added{The constraint} $W^0(x)=0$ \added{commutes with all other constraints and is therefore a first-class constraint. For this reason, we must add a gauge-fixing constraint. Continuing, the number of remaining constraints will be odd. Thus, }  \deleted{and} \added {along with the constraint} $\partial_{i}F^{0i}(x)+J^0(x)=0,$ \deleted{are first-class constraints} \added{we must add another gauge-fixing constraint to ensure a non-zero determinat of the constraint Poisson brakets.} \deleted{and} \added{In this way,}we need to consider \deleted{two} \added{these} additional constraints alongside \deleted{them} \added{the two constraints thet have been raised} :
\be\label{4.18c}
v_{0}(x)=0, ~~~ \partial^{i}\Pi_{0i}(x)=0.
\ee
\added{Considering the primary constraint $\Pi_{0i}+v_{i}=0$, we can write the second gauge-fixing constraint as $\partial^{i}v_{i}=0$.}
Substituting these constraints into equation \eqref{2.43} \deleted{and simplifying},\added{and noting that the determinant in the measure is a constant,} the path integral takes the following form:

\bea\label{4.19} 
\int\mathcal{D}F^{\mu<\nu}\delta(\partial_{\mu}F^{\mu0}(x)-J^{0}(x))e^{-\frac{i}{4}\int d^4xF_{\mu\nu}(x)F^{\mu\nu}(x)}~~~~~~~~~~~~~~~~~~~~~~~~~~~~\nonumber\\
\times\int\mathcal{D}v_{i}e^{-i\int d^4x~ v_{i}(x)(\partial_{\mu}F^{\mu i}(x)-J^i(x))}\added{\delta(\partial^iv_{i})}\delta(F^{ij}(x)-\partial^iv^j(x)+\partial^j v^i(x))\deleted{.}\added{,}
\eea
\deleted{Therefore out of the four constraint equations in \eqref{4.16a}, only one is transformed into a delta functional.}
\added{which, of course, is only one of the constraints in \eqref{4.16a}, transformed into the delta functional.} 
\added{Continuing, we can write the partition function:}
\bea\label{4.19a}
\added{\int\mathcal{D}F^{0i}\delta(\partial_{i}F^{0i}(x)+J^{0}(x))e^{-\frac{i}{2}\int d^4xF_{0i}(x)F^{0i}(x)}}~~~~~~~~~~~~~~~~~~~~~~~~~~~~~~~~~~~~~~~\nonumber\\
\added{\times\int\mathcal{D}v_{i}\delta(\partial^iv_{i})e^{-i\int d^4x v_{i}(x)(\partial_{0}F^{0 i}(x)+\nabla^2v^i(x)-J^i(x))})e^{-\frac{i}{2}\int d^4x ~\partial_{i}v_{j}(x)(\partial^iv^j(x)-\partial^jv^i(x))}}
\eea
\added{Finally, we can consider the electric field as a classical degree of freedom that applies to the Coloumb equation, while the vector potential will be a quantum degree of freedom.}

\section{Discussion}
The imposition of external constraints on a quantum system effectively treats \deleted{some of the system's variables} \added{certain variables} as classical and measurable. For example, equation \eqref{3.19} indicates that $c_{\alpha}$ and $d_{\alpha}$ \deleted{are} \added{behave as} classical variables \deleted{that are measured to have the}\added{, constrained to take the measured} value of zero.  If the Hamiltonian is non-singular,
equation \eqref{2.8} can \deleted{also} be interpreted in the \added{same} manner \deleted{described}.

In equation \eqref{3.8}, where $b_{\alpha i}\neq0$, \deleted{the measurement of}\added{the condition} $\phi_{\alpha}(q,p,v)=0 $ \deleted{is conducted in the presence of a constraint force.}\added{ is interpreted as a measurement performed in the presence of a constraint force.}
\deleted{It is as if the fields are quantized}\added{ This setup is analogous to quantizing fields} in a dynamical classical field background. For \deleted{example}\added{ instance}, the SO(3,1) quantum gauge theory is \deleted{based on} \added{formulated on the basis of} classical diffeomorphism\deleted{s}\added{ symmetry}\cite{wies, wiesx}.

If we want to ensure that there is no constraint force, the variables $v_{\alpha}$ and $w_{\alpha}$\deleted{should} \added{must} be removed from equations \eqref{3.6}, \eqref{3.7}, and \eqref{3.8}.
Consequently, the functional integral becomes:
\be\label{6.1}
\int\mathcal{D}q\mathcal{D}p (\det\deleted{\cancel{[}}\added{\{}\phi_{\alpha},\phi_{\beta}\deleted{\cancel{]}}\added{\}})^{\frac{1}{2}}\delta(\phi_{\alpha}(q,p))e^{ i\int dt \;\dot{q}_{i}p_{i}-H(q,p)},
\ee
where
\be\label{6.2}
\phi_{\alpha}(q,p)=b_{\alpha i}(q)A^{-1}_{ij}(q)(p_{j}-B_{j}(q))+c_{\alpha}(q)=0, 
\ee
which are measured at all times. This imposes the necessary condition on the Hamiltonian:
\be\label{6.2a}
\{H(q,p),\phi_{\alpha}(q,p)\}=0.
\ee

So far, the classical quantities have been assumed to take the value zero. Now, let us consider the case where there are no constraint equations, and we assume that some canonical duals are classical,
with various values being measured. In this case, we \deleted{write}\added{express} the functional integral as follows:

\bea\label{6.3}
U(q^{*}_{a},p^{*}_{a})=\displaystyle{\lim_{T\to \infty(1+i\epsilon)}}<q^{*}_{A}(-T)|e^{i2TH^{*}(q^{*},p^{*})}|q^{*}_{A}(T)>\nonumber\\
=\int \mathcal{D}q^{*}_{A}\mathcal{D}p^{*}_{A}e^{i\int dt \dot{q}^{*}_{A}p^{*}_{A}-H^{*}(q^{*}_{a},p^{*}_{a},q^{*}_{A},p^{*}_{A})},
\eea
where $\eta^{*}=(q^{*}_{a},p^{*}_{a})$ are classical fields and $(q^{*}_{A},p^{*}_{A})$ are quantum fields.
If we consider the classical fields \deleted{to reperesent}\added{as representing} the background state of the quantum system, \deleted{then}the following necessary conditions must hold between the measured values at all times and the Hamiltonian:
\be\label{6.4}
 \frac{d\eta^{*}}{dt}=\frac{\int \mathcal{D}q^{*}_{A}\mathcal{D}p^{*}_{A}\{\eta^{*} , H^{*}\} e^{i\int dt \dot{q}^{*}_{A}p^{*}_{A}-H^{*}}}{\int \mathcal{D}q^{*}_{A}\mathcal{D}p^{*}_{A}e^{i\int dt \dot{q}^{*}_{A}p^{*}_{A}-H^{*}}}.
\ee
Otherwise, the introduction of a constraint force and a modified Hamiltonian becomes necessary.
Generally, the coexistence of classical and quantum quantities \deleted{is employed}\added{arises} when undetermined fluctuations are superimposed on a determined background.

\end{document}